\documentclass[11pt, oneside]{article}   	% use "amsart" instead of "article" for AMSLaTeX format
\usepackage{geometry}                		% See geometry.pdf to learn the layout options. There are lots.
\usepackage{graphicx}				% Use pdf, png, jpg, or eps with pdflatex; use eps in DVI mode
\usepackage{amssymb}
\usepackage{amsmath}
\usepackage{subfig}
\usepackage{caption}
\usepackage{comment}
\usepackage{amsthm,verbatim,amsfonts,amscd,float,physics}
\usepackage{appendix}

\usepackage[utf8]{inputenc}
\usepackage{xcolor}
\numberwithin{equation}{section}
\begin{document}
\title{Worldline Path Integrals for \\
Gauge Fields and Quantum Computing }
\author{Yuan Feng$^{(1)}$, Michael McGuigan $^{(2)}$\\
(1) Pasadena City College,  (2) Brookhaven National Laboratory\\ 
email: michael.d.mcguigan@gmail.com}
\date{}
\maketitle
\begin{abstract}
We study different aspects the worldline path integrals with gauge fields using quantum computing. We use the Variational Quantum Eigensolver (VQE) and Evolution of Hamiltonian (EOH) quantum algorithms and IBM QISKit to perform our computations. We apply these methods to the path integral of a particle moving in a Abelian and non-Abelian  background gauge field  associated with a constant magnetic field and the field of a chromo-magnetic field.  In all cases we found excellent agreement with the classical computation. We also discuss the insertion of  vertex operators into the worldline path integrals to study scattering and show how to represent them  using unitary operators and quantum gates on near term quantum computers.
\end{abstract}
\newpage

\section{Introduction}

Recently there has been a great deal of interest in the quantum simulation of non-Abelian gauge fields. Most of these studies are restricted to lower dimensions because of the large amount of qubits and large depth quantum circuits that are involved in performing a quantum field simulation. In this paper we explore  quantum simulation of gauge fields using the worldline formalism. We study the quantum simulation of a particle moving in a constant magnetic field in both Cartesian and polar coordinates. We also  study the world line Hamiltonian for a non-Abelian gauge field and discuss the use of vertex operators for the quantum simulation of nonlinear theories. We use the Variational Quantum Eigensolver (VQE) quantum algorithm and Evolution of Hamiltonian  (EOH) quantum algorithm to obtain our results and obtain excellent agreement with classical computations.

\section{Quantum computer simulation for a constant magnetic field}

One example of a worldline path integral in particle propagation in a constant electromagnetic field. For a constant magnetic field the non-relativistic Lagrangian is given by \cite{Seri}\cite{Cheng}\cite{Feynman}:
\begin{equation}L = \frac{m}{2}({{\dot x}^2} + {{\dot y}^2} + {{\dot z}^2}) + \frac{{eB}}{{2c}}(x\dot y - \dot x y)\end{equation}
Where $B$ is the magnitude of the magnetic field in the $z$ direction.
This is written in terms of the vector potential as:
\begin{equation}L = \frac{m}{2}({{\dot x}^2} + {{\dot y}^2} + {{\dot z}^2}) + {A_x}(x,y,z)\dot x + {A_y}(x,y,z)\dot y + {A_z}(x,y,z)\dot z\end{equation}
and with vector potential given by:
\begin{equation}A = ( - \frac{{eB}}{{2c}}y,\frac{{eB}}{{2c}}x,0)\end{equation}
one has the Hamiltonian:
\begin{equation}H = \frac{1}{{2m}}{(\vec p - \frac{q}{c}\vec A(x))^2}\end{equation}
which can be written:
\begin{equation}H = \frac{1}{2}{\left( {{p_x} + \frac{B}{2}y} \right)^2} + \frac{1}{2}{\left( {{p_y} - \frac{B}{2}x} \right)^2}\end{equation}
for a constant magnetic field the energy eigenvalues are:
\begin{equation}{E_n} = \left| B \right|(n + \frac{1}{2})\end{equation}
with ground state wave function:
\begin{equation}{\psi _0}(x,y) = \sqrt {\frac{{\left| B \right|}}{{2\pi }}} {e^{ - \frac{{\left| B \right|}}{4}\left( {{x^2} + {y^2}} \right)}}\end{equation}
Expanding out the terms in $B$ the Hamiltonian can be written:
\begin{equation}H = \frac{1}{2}(p_x^2 + p_y^2) + \frac{1}{2}{\left( {\frac{B}{2}} \right)^2}({x^2} + {y^2}) - \frac{B}{2}(x{p_y} - y{p_x})\end{equation}
which shows the relation to the two dimensional isotropic simple harmonic oscillator.
The Kernel function 
\begin{equation}K({x_i},{y_i},{x_f},{y_f};t) = \left\langle {{x_f},{y_f}} \right|{e^{ - iHt}}\left| {{x_i},{y_i}} \right\rangle \end{equation}
can be evaluated using path integrals and in Cartesian coordinates is given by:
\begin{equation}K = {\left( {\frac{1}{{2\pi T}}} \right)^{3/2}}\left( {\frac{{BT/2}}{{\sin (BT/2)}}} \right){e^{i\frac{{{{({z_f} - {z_i})}^2}}}{{2T}}}}\exp \left[ {\frac{{iBT/2}}{{\tan (BT/2)}}\left( {{{({x_f} - {x_i})}^2} + {{({y_f} - {y_i})}^2} + B({y_f}{x_i} - {x_f}{y_i})} \right)} \right]\end{equation}
One can also use polar coordinates to study the particle in a constant magnetic field. In terms of polar coordinates the Hamiltonian is:
\begin{equation}H = \frac{1}{2}{\rho ^{ - 1/2}}{p_\rho }\rho {p_\rho }{\rho ^{ - 1/2}} + \frac{1}{2}{\left( {\frac{B}{2}} \right)^2}{\rho ^2} + \frac{1}{2}\frac{{{L_z^2}}}{{{\rho ^2}}} - \frac{B}{2}L_z\end{equation}
and the eigenvalues and eigenfunctions are give by
\begin{equation}{E_{n,m}} = (n + 1 - m)\frac{B}{2}\end{equation}
and 
\begin{equation}{\psi _{n,m}}(`\rho ,\varphi ,t) = {\left[ {\frac{{n!\left( {B/2} \right)}}{{\pi \left( {n + \left| m \right|} \right)!}}} \right]^{1/2}}{\left[ {{{\left( {\frac{B}{2}} \right)}^{1/2}}\rho } \right]^{\left| m \right|}}{e^{ - \frac{B}{4}{\rho ^2}}}L_n^{\left| m \right|}\left( {\frac{B}{2}{\rho ^2}} \right){e^{im\varphi }}{e^{i{E_{n.m}}t}}\end{equation}
where $L_n^\alpha$ are generalized Laguerre polynomials.
In terms of polar coordinates the K function is:
\begin{equation}K = \left( {\frac{1}{{2\pi i}}} \right)\frac{{B/2}}{{\sin (BT/2)}}{e^{ - \frac{B}{4}\left( {\rho _f^2 + \rho _i^2} \right)}}{e^{i\frac{B}{4}\left( {\rho _f^2 + \rho _i^2} \right)\frac{{{e^{ - iBT/2}}}}{{\sin (BT/2)}}}}\sum\limits_{m =  - \infty }^\infty  {{e^{im({\varphi _f} - \varphi_i  + \frac{B}{2}{T})}}{I_{\left| m \right|}}\left[ { - i\frac{B}{2}{\rho _f}{\rho _i}\frac{1}{{\sin \left( {BT/2} \right)}}} \right]} \end{equation}
where $I_\nu (z)$ is a modified Bessel function. Having reviewed the energies and Kernel functions for motion of particle in a constant magnetic field we now turn the the quantum computation of these quantities in terms of qubits and quantum circuits.

\subsection{Quantum computation with the Variational Quantum Eigensolver}

The Variational Quantum Eigensolver (VQE) is a hybrid classical quantum algorithm that can be run on near term quantum hardware because the circuit depth can be kept small for the algorithm, at least for simple variational wave forms \cite{Kandala}. Like the traditional variational method one optimizes
\begin{equation}{E_0}({\theta _i}) = \frac{{\left\langle {\psi ({\theta _i})} \right|H\left| {\psi ({\theta _i})} \right\rangle }}{{\left\langle {\psi ({\theta _i})} \right|\left. {\psi ({\theta _i})} \right\rangle }}\end{equation}
with variational parameters $\theta_i$ to obtain an upper bound on the value of the ground state energy of a Hamiltonian. One uses an optimizer to determine the minimum of the energy expectation value and the expectation value is evaluated on the quantum computer. QISKit provides several optimizers. In this paper we use the The the  statevector simulator with the  Sequential Least SQuares Programming (SLSQP) optimizer. For the VQE the variational wave functions are represented as unitary matrices acting on qubits through quantum gates and the the parameters are rotation angles associated with these gates. 

For the Hamiltonian (2.5) 
\begin{equation}H = \frac{1}{2}{\left( {{p_x} + \frac{B}{2}y} \right)^2} + \frac{1}{2}{\left( {{p_y} - \frac{B}{2}x} \right)^2}\end{equation}
and using the Harmonic oscillator basis
to map the the Hamiltonian  in terms of qubits we use
\begin{equation} 
 {Q} = \frac{1}{\sqrt{2}}\begin{bmatrix}
 
   0 & {\sqrt 1 } & 0 &  \cdots  & 0  \\ 
   {\sqrt 1 } & 0 & {\sqrt 2 } &  \cdots  & 0  \\ 
   0 & {\sqrt 2 } &  \ddots  &  \ddots  & 0  \\ 
   0 & 0 &  \ddots  & 0 & {\sqrt {N-1} }  \\ 
   0 & 0 &  \cdots  & {\sqrt {N-1} } & 0  \\ 
\end{bmatrix}
  \end{equation}
while for the momentum operator we have:
\begin{equation}
 P = \frac{i}{\sqrt{2}}\begin{bmatrix}
 
   0 & -{\sqrt 1 } & 0 &  \cdots  & 0  \\ 
   {\sqrt 1 } & 0 & -{\sqrt 2 } &  \cdots  & 0  \\ 
   0 & {\sqrt 2 } &  \ddots  &  \ddots  & 0  \\ 
   0 & 0 &  \ddots  & 0 & -{\sqrt {N-1} }  \\ 
   0 & 0 &  \cdots  & {\sqrt {N-1} } & 0  \\ 
\end{bmatrix}
  \end{equation}
taking $N=16$ we can represent the operators using tensor products with each factor in the tensor products as below:
\begin{equation}\begin{array}{l}
x = Q \otimes {I_{16}}\\
y = {I_{16}} \otimes Q\\
{p_x} = P \otimes {I_{16}}\\
{p_y} = {I_{16}} \otimes P
\end{array}\end{equation}
We obtain the result for the$E_0 $ from the  VQE for the above two boson Hamiltonian is osc basis where each boson is represents by a $16 \times 16$ matrix the total Hamiltonian uses a $256 \time 256 $ matrix and is represented using eight qubits. The results are shown in figure 1 and table 1 
and agree  well with the exact value.
\begin{figure}
\centering
\minipage{0.5\textwidth}
  \includegraphics[width=\linewidth]{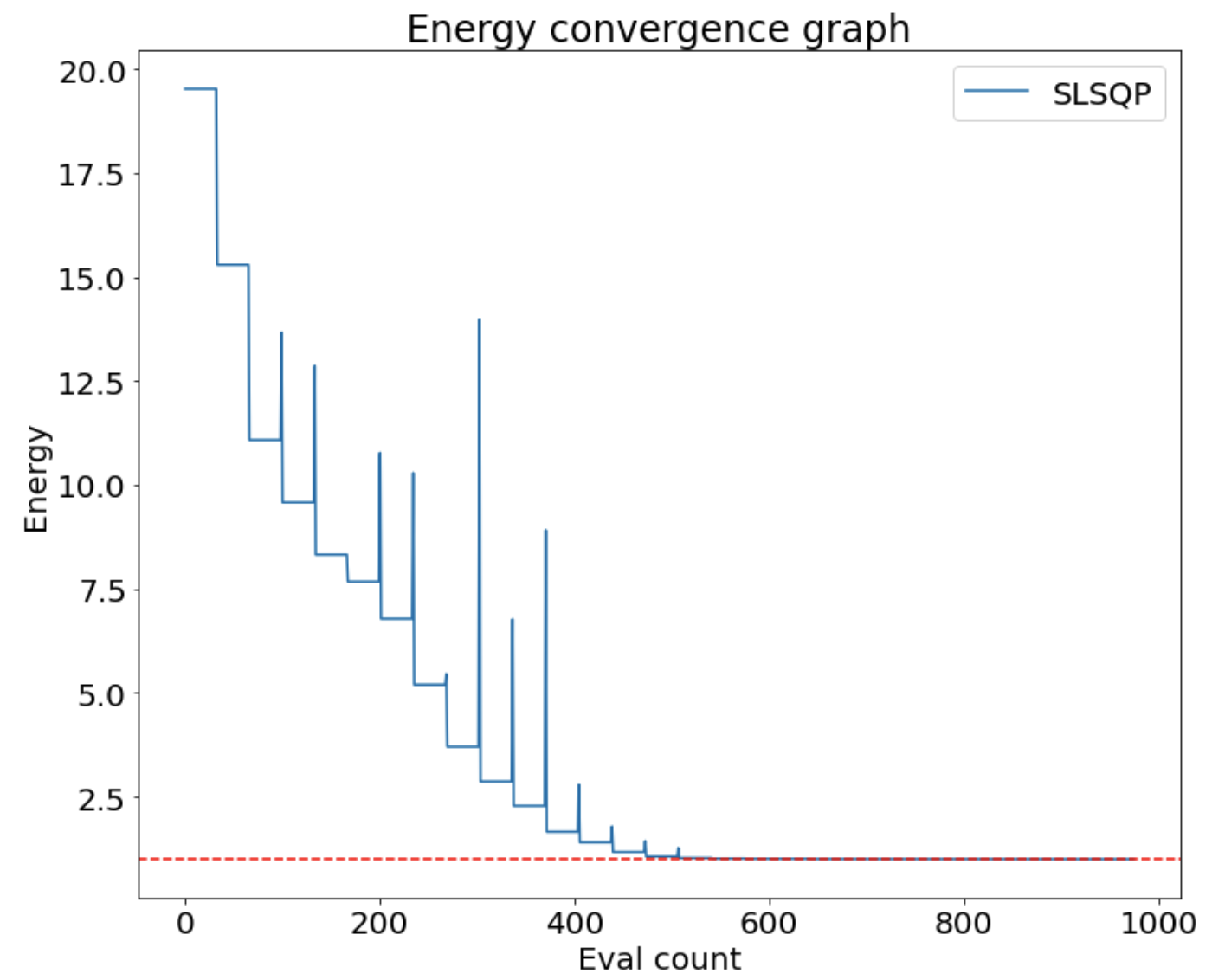}
%\label{fig:awesome_image1}
\endminipage\hfill
\minipage{0.5\textwidth}
  \includegraphics[width=\linewidth]{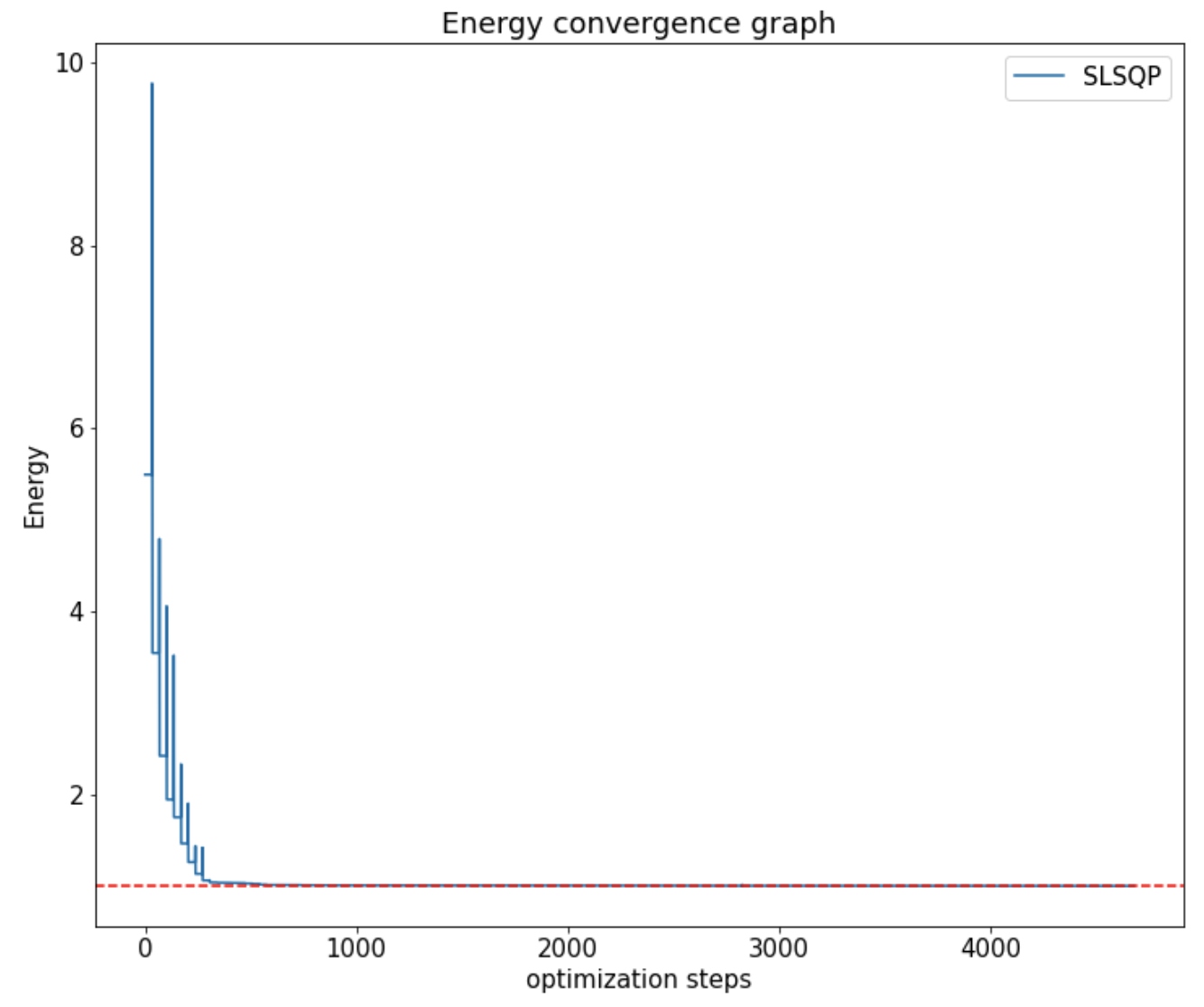}
\endminipage\hfill
\caption{(Left) VQE computation of ground state energy of a particle moving in a constant magnetic field using Cartesian coordinates and the SLSQP optimizer (right) VQE computation of ground state energy of a particle moving in a constant magnetic field using polar coordinates and  also using the SLSQP optimizer. }
\end{figure}

\begin{table}[ht]
\centering
\begin{tabular}{|l|l|l|l|}
\hline
Hamiltonian       & No. of Qubits  & Exact Result & VQE Result \\ \hline
%Position   &    \(H_-\)         & -0.0004   & 42 \\ \hline
%Position   &    \(H_+\)         & 4.4342   & 42 \\ \hline
%Finite Difference   & \(H_-\) & 0.2111   & 38 \\ \hline
%Finite Difference   & \(H_+\) & 4.4643  & 38 \\ \hline
Constant Magnetic Cartesian & 8  & 1.0 & 1.0000003553  \\ \hline
Constant Magnetic Polar  & 4  & 0.9980452 & 0.99859198   \\ \hline

\end{tabular}
\caption{\label{tab:BasisCompare}  VQE results fora particle moving in a constant magnetic field using Cartesian and Polar coordinates.  and using the oscillator basis. The Hamiltonian were mapped to 8-qubit operators for Cartesian coordinates and 4-qubit operators for Polar coordinates. The quantum circuit for each simulation utilized an \(R_y\) variational form, with a fully entangled circuit of depth 3. The backend used was a statevector simulator. The Sequential Least SQuares Programming (SLSQP) optimizer was used, with a maximum of 600 iterations.}
\end{table}

For the polar coordinates form of the Hamiltonian 
\begin{equation}H = \frac{1}{2}{\rho ^{ - 1/2}}{p_\rho }\rho {p_\rho }{\rho ^{ - 1/2}} + \frac{1}{2}{\left( {\frac{B}{2}} \right)^2}{\rho ^2} + \frac{1}{2}\frac{{{L_z^2}}}{{{\rho ^2}}} - \frac{B}{2}L_z\end{equation}
In this case we don't need a tensor product to contruct the Hamiltonian and 4 qubits suffice for the quantum computation. The results of the VQE computations are shown in figure 1 and table 2 and are in excellent agreement with the classical computations. This is possible because of the similarity of the particle in a constant magnetic field and the two dimensional harmonic oscillator for which the variational forms in QISKit has a strong overlap with the ground state wave function.
\begin{figure}
\centering
\minipage{.7\textwidth}
  \includegraphics[width=\linewidth]{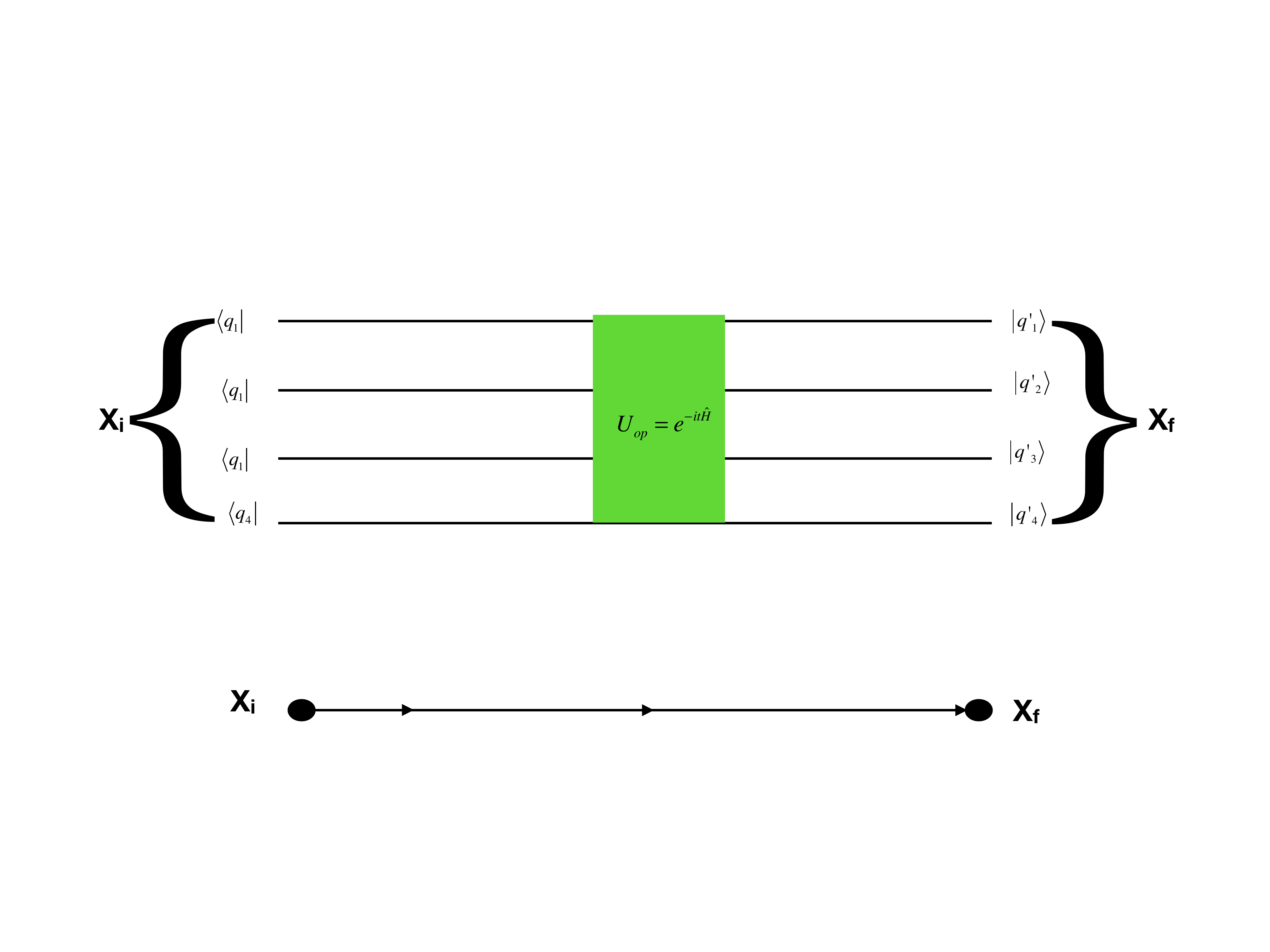}
%\label{fig:awesome_image1}
\endminipage\hfill
\caption{(top) Evolution of Hamiltonian for particle moving in constant magnetic field  in terms of a quantum circuit where the green rectangle indicated the unitary operator for time evolution (bottom) Feynman propagator of Kernel function in terms of a Feynman graph corresponding to the top circuit. }
\end{figure}
\begin{figure}
\centering
\minipage{0.3\textwidth}
  \includegraphics[width=\linewidth]{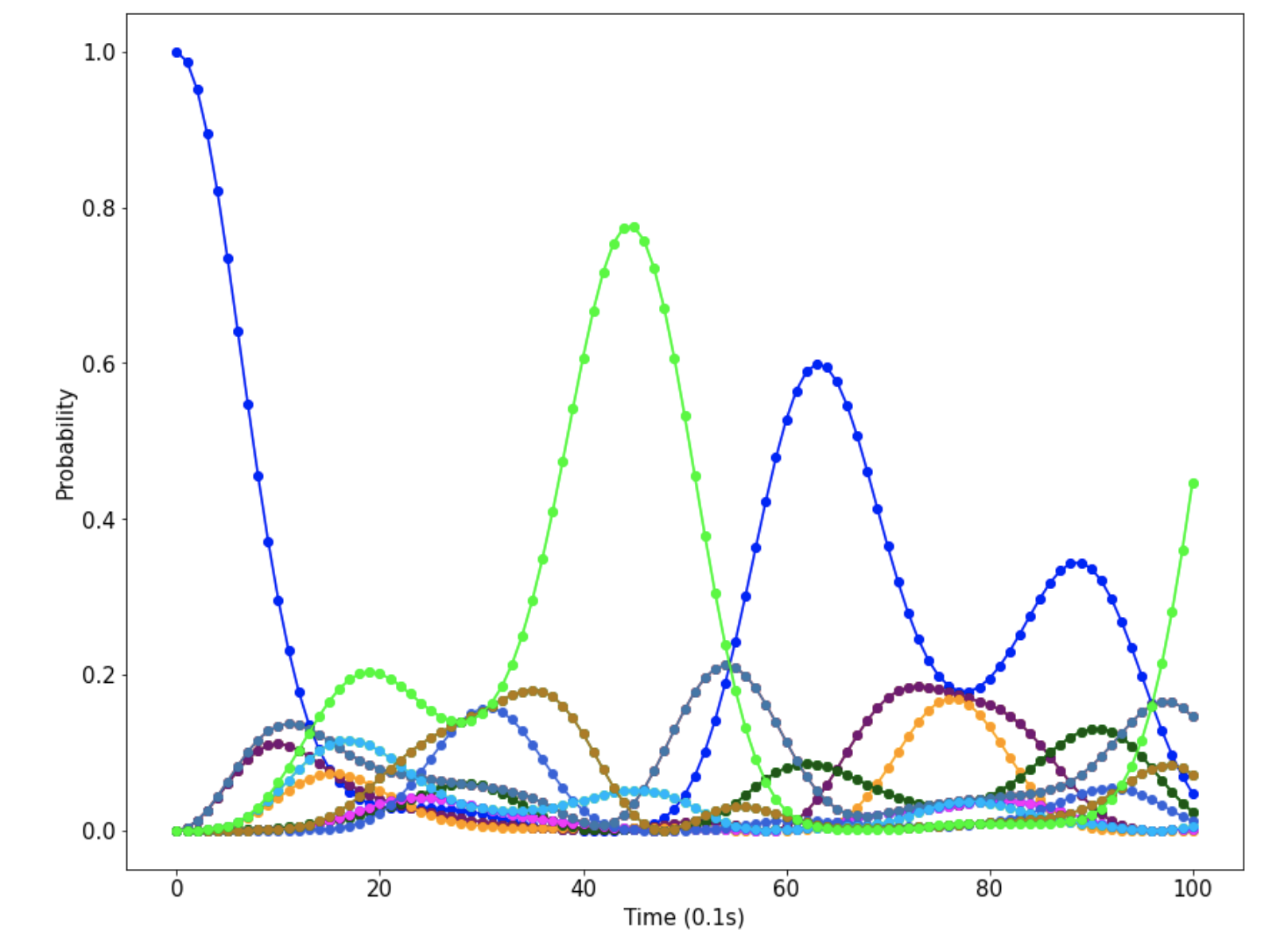}
%\label{fig:awesome_image1}
\endminipage\hfill
\minipage{0.3\textwidth}
  \includegraphics[width=\linewidth]{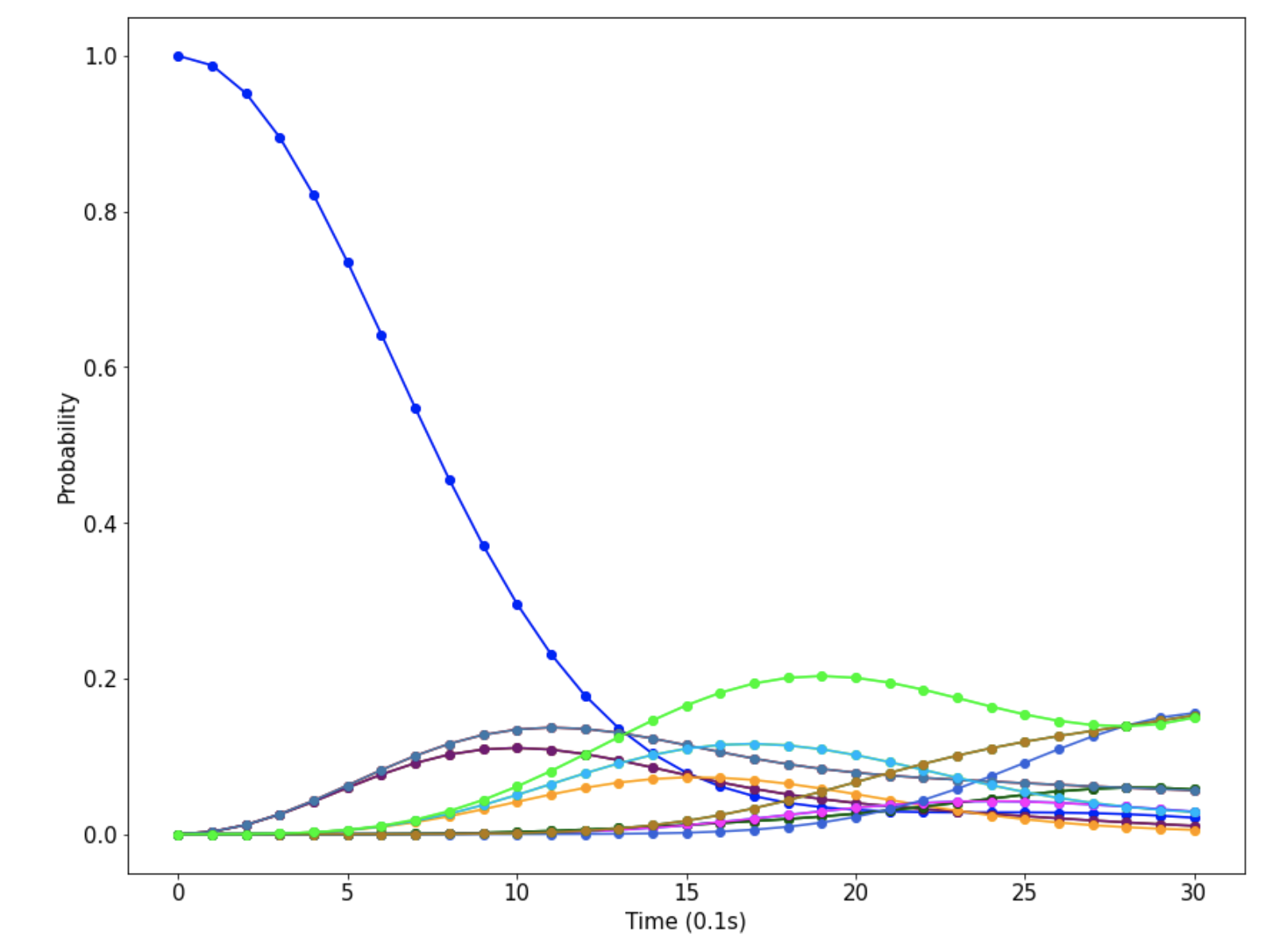}
\endminipage\hfill
\minipage{0.3\textwidth}
  \includegraphics[width=\linewidth]{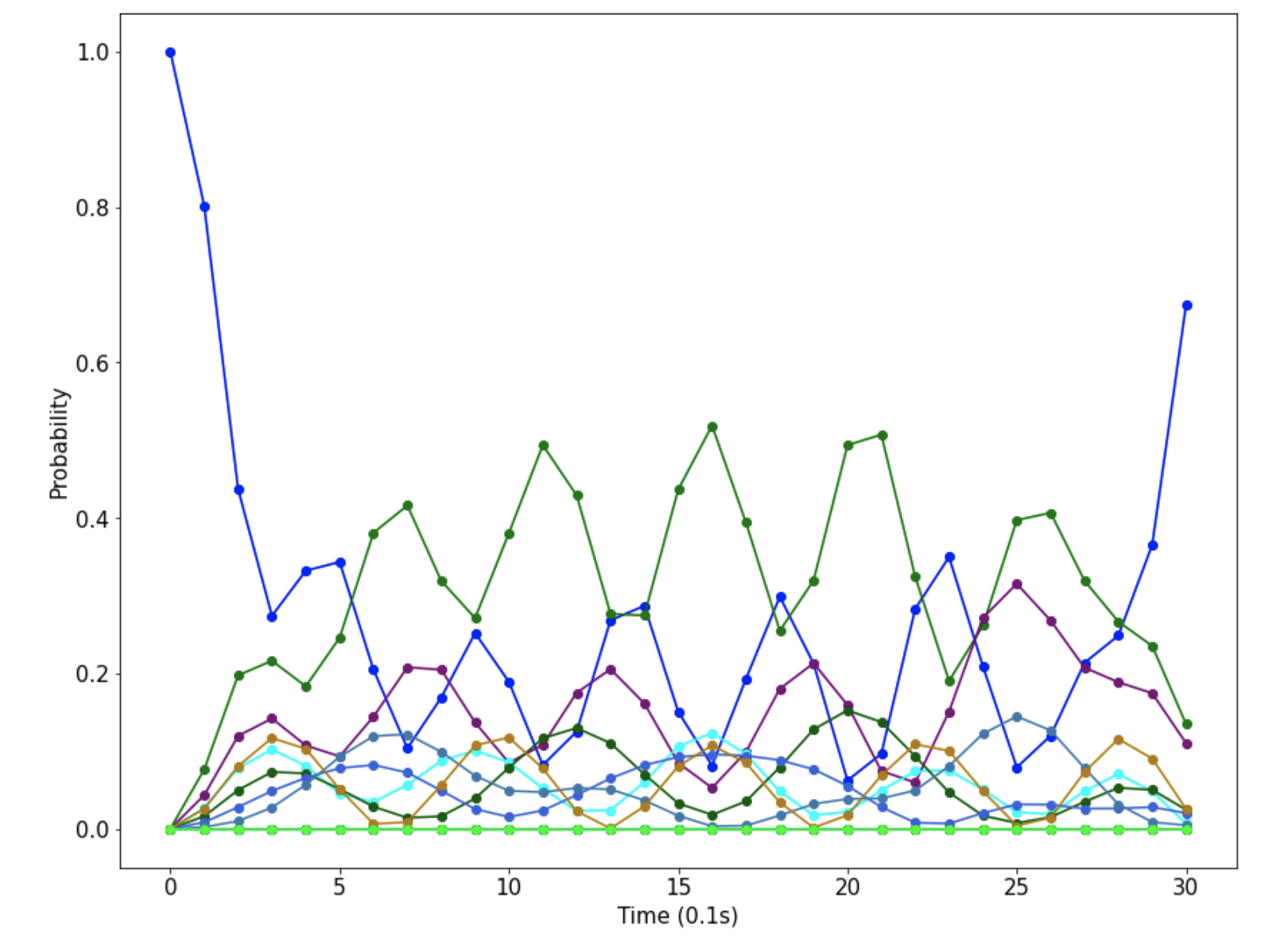}
\endminipage\hfill
\caption{(Left) Results of EOH quantum computation  for particle moving in constant magnetic field in Cartesian coordinates (middle) closeup view of the EOH for Cartesian coordinates ( right ) Results of EOH quantum computation  for particle moving in constant magnetic field in polar coordinates .}
\end{figure}

\subsection{Quantum computation with the Evolution of Hamiltonian algorithm.}

A second quantum algorithm we applied to worldline path integrals of gauge fields is the Evolution of Hamiltonian algorithm. The Evolution of Hamiltonian (EOH) algorithm evaluates 
\begin{equation}K({X _i},X _f;T) = \left\langle {\psi ({X _f})} \right|{e^{ - i t H }}\left| {\psi ({X _i})} \right\rangle \end{equation}
the transition amplitude between an initial state  and a final state  with $H$ the Hamiltonian and $t$ the time of the evolution and is depicted in figure 2.
When applying the EOH algorithm it is convenient to use the position basis in which the position operator is diagonal instead of the oscillator basis described above. In the position basis the momentum matrix is dense and constructed from the position operator using a Sylvester matrix $F$. In the position basis the position matrix is:
\begin{equation}
{\left( {{Q_{pos}}} \right)_{j,k}} = \sqrt {\frac{{2\pi }}{{4N}}} (2j - (N + 1)){\delta _{j,k}}
\end{equation}
and the momentum matrix is:
\begin{equation}{P_{pos}} = {F^\dag }{Q_{pos}}F\end{equation}
where 
\begin{equation}{F_{j,k}} = \frac{1}{{\sqrt N }}{e^{\frac{{2\pi i}}{{4N}}(2j - (N + 1))(2k - (N + 1))}}\end{equation}
Then the position and momentum operators are defined similarly to the above in terms of tensor products with $Q$ replaced by $Q_{pos}$ and $P$ replaced by $P_{pos}$. The results of the EOH computation using the Trotter-Suzuki approximation in QISKit of the particle in a constant magnetic field are show in figure 3. The multiple curves in the graphs indicate the various transition probabilities from initial position $(x_i,y_i)=(0,0) $ to a final position $(x_f,y_f)$
and are  in excellent agreement with the classical results obtained by directly exponentiating the Hamiltonian matrix.

\section{Quantum computer simulation for a non-Abelian gauge field}

The quantum computer simulation of a particle in a background non-Abelian gauge field is  more complicated than the quantum simulation of a particle in an Abelian gauge field discussed above. This is because the part of the Lagrangian involving the non-Abelian gauge field has to be path ordered for gauge invariance and this is difficult to implement directly. Instead one introduces worldline fermions into the 0+1 dimensional path integral associated with the non-Abelian symmetry group to implement the path ordering \cite{Pisarski:2006yk}\cite{KorthalsAltes:1994xx}\cite{Mueller:2017wom}\cite{Mueller:2020vha}
\cite{Callan:1989nz}
\cite{Fan:2017uqy}
\cite{Sen:1985eb}
\cite{deAlwis:1986rw}
\cite{Ellwanger:1988cc}. Then one can treat the systems as a more traditional Lagrangian system and can implement quantum simulation more straightforwardly.
The Lagrangian of a particle moving in a non-Abelian gauge field then is given by
\begin{equation}L = \frac{1}{2}{{\dot x}^{i2}} + {\psi ^a}{{\dot \psi }^a} + {f_{abc}}A_i^a(x){{\dot x}^i}{\psi ^b}{\psi ^c}\end{equation}
where $\psi^a$ are world line fermion fields which take into account the path ordering and $f_{abc}$ are the structure constant of the Lie group. The Hamiltonian is then:
\begin{equation}H = \frac{1}{2}{\left( {{p_i} + {f_{abc}}A_i^a(x){\psi ^b}{\psi ^c}} \right)^2}\end{equation}
In a background field expansion about a reference point $x_0$ 
%\begin{equation}H = \frac{1}{2}{\left( {{p_x} + {f_{abc}}F_{xy}^a({x_0})y{\psi ^b}{\psi ^c}} \right)^2} + \frac{1}{2}{\left( {{p_y} - {f_{abc}}F_{xy}^a({x_0})x{\psi ^b}{\psi ^c}} \right)^2}\end{equation}
and  a background field expansion using normal coordinates the Hamiltonian can be written in terms of the field strength as:
\begin{equation}H = \frac{1}{2}{\left( {{p_i} + {f_{abc}}F_{ij}^a({x_0}){x^j}{\psi ^b}{\psi ^c}} \right)^2}\end{equation}
a form similar to the Abelian case discussed above. 

To proceed further one needs a solution to the non-Abelian backgorund gauge field equations of motion or the Yang-Mills equation \cite{Marinho:2009tm}
\cite{Norfjand:2019gof}
\cite{Marciano:1977sm}
\cite{Actor:1979in}
\cite{Moody:1985ty}. For the non-Abelian $SU(2)$ gauge field an ansatz  used for the Wu-Yang is given by:
\begin{equation}A_i^a(x) = {\varepsilon _{iab}}\frac{{{x^b}}}{{{r^2}}}f(r)\end{equation}
defining $g=f-1$ the equations of motion reduce to:
\begin{equation}g'' = \frac{g}{{{r^2}}}({g^2} - 1)\end{equation}
which for small r has the expansion
\begin{equation}g \approx 1 - {r^2} + \frac{3}{{10}}{r^4} +  \ldots \end{equation}
and large $r$ we have:
\begin{equation}g \approx 1 - \frac{1}{r} + \frac{3}{4}\frac{1}{{{r^2}}} +  \ldots \end{equation}
These expressions can be used to study the quantum simulation of a particle in a background non-Abelian gauge field but other forms can be studied as well depending on the application. 

\subsection{Quantum computation with the VQE for Hamiltonian for a non-Abelian gauge field}

To map the the Hamiltonian for a particle moving in a non-Abelian gauge field in terms of qubits we use $P$ and $Q$ from (2.17) and (2.18) with 
 $N=4$ and form the tensor products:
\begin{equation}\begin{array}{l}
x = Q \otimes {I_4} \otimes {I_4} \otimes {I_8}\\
y = {I_4} \otimes Q \otimes {I_4} \otimes {I_8}\\
z = {I_4} \otimes {I_4} \otimes Q \otimes {I_8}\\
{p_x} = P \otimes {I_4} \otimes {I_4} \otimes {I_8}\\
{p_y} = {I_4} \otimes P \otimes {I_4} \otimes {I_8}\\
{p_z} = {I_4} \otimes {I_4} \otimes P \otimes {I_8}
\end{array}\end{equation}
For $SU(2)$ we have three worldline fermions which are represented by
\begin{equation}\begin{array}{l}
{\psi _1} = {I_{64}} \otimes \left( {\begin{array}{*{20}{c}}
0&1\\
0&0
\end{array}} \right) \otimes {I_2} \otimes {I_2}\\
{\psi _2} = {I_{64}} \otimes {I_2} \otimes \left( {\begin{array}{*{20}{c}}
0&1\\
0&0
\end{array}} \right) \otimes {I_2}\\
{\psi _3} = {I_{64}} \otimes {I_2} \otimes {I_2} \otimes \left( {\begin{array}{*{20}{c}}
0&1\\
0&0
\end{array}} \right)
\end{array}\end{equation}
For the quantum computation for non-Abelian gauge field we use the Hamilton  
\begin{equation}H = \frac{1}{2}{\left( {{p_x} + B( - y{\psi ^1}{\psi ^2} + z{\psi ^3}{\psi ^1}} \right)^2} + \frac{1}{2}{\left( {{p_y} + B( - z{\psi ^2}{\psi ^3} + x{\psi ^1}{\psi ^2}} \right)^2} + \frac{1}{2}{\left( {{p_z} + B( - x{\psi ^3}{\psi ^1} + y{\psi ^2}{\psi ^3}} \right)^2}\end{equation}
where $ B = -\frac{{ g_m}}{{{r^2}}}$.

For three bosons $x_i$ represented as $4\times 4$ so six qubits are used to represent the bosons and for three fermions $\psi_i$ we use three  qubits. The full Hamiltonian for a particle moving in an $SU(2)$ gauge field uses nine qubits and is represented by a $512 \times 512$ matrix. The results using the VQE algorithm and the SLSQP optimizer are shown in figure 4 and table 2. The results are not as accurate as for the Abelian gauge field. One possible reason is that the variational form used in the optimization is parametrized by angles associated with quantum gates does not overlap as strongly with the ground state for non-Abelian backgrounds as it does in the Abelian case.
\begin{figure}
\centering
\minipage{0.5\textwidth}
  \includegraphics[width=\linewidth]{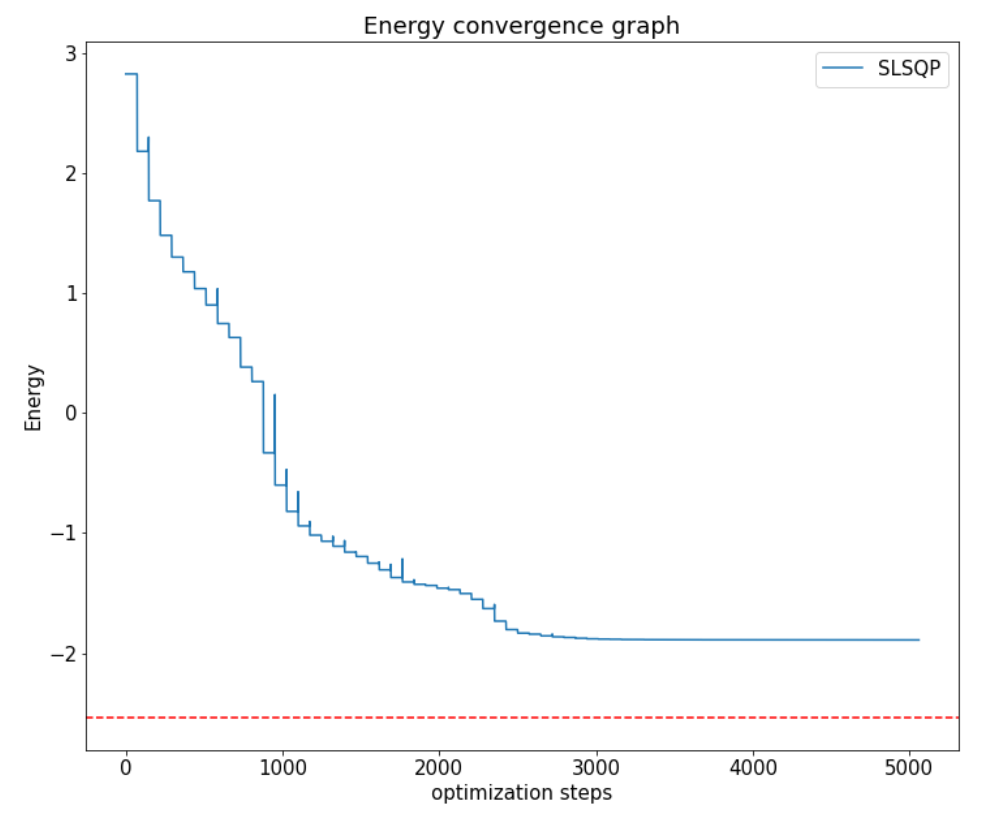}
%\label{fig:awesome_image1}
\endminipage\hfill
\minipage{0.5\textwidth}
  \includegraphics[width=\linewidth]{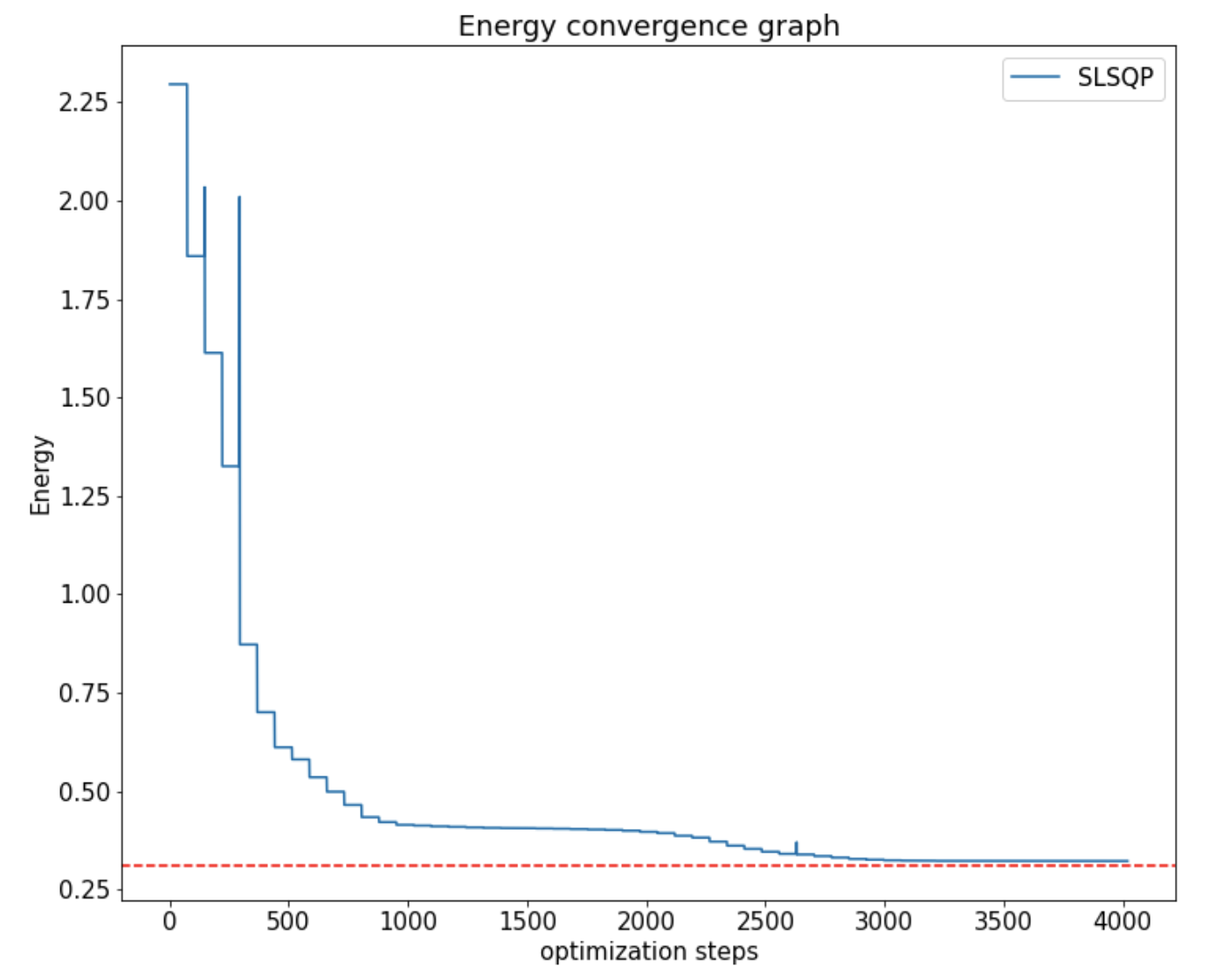}
\endminipage\hfill
\caption{(Left) VQE computation of ground state energy of a particle moving in a non-Abelian gauge field  with $g_m = 2$ and the SLSQP optimizer (right) VQE computation of ground state energy of a particle moving in a non-Abelian gauge field  with $g_m = .2$ and  also using the SLSQP optimizer. }
\end{figure}
\begin{table}[ht]
\centering
\begin{tabular}{|l|l|l|l|}
\hline
Hamiltonian       & No. of Qubits  & Exact Result & VQE Result \\ \hline
%Position   &    \(H_-\)         & -0.0004   & 42 \\ \hline
%Position   &    \(H_+\)         & 4.4342   & 42 \\ \hline
%Finite Difference   & \(H_-\) & 0.2111   & 38 \\ \hline
%Finite Difference   & \(H_+\) & 4.4643  & 38 \\ \hline
 Magnetic Monopole $g_m = 2$ & 9  & -2.53854786 & -1.89074415 \\ \hline
Magnetic Monopole $g_m = .2$  & 9  & 0.31120022 &   0.32279234  \\ \hline

\end{tabular}
\caption{\label{tab:BasisCompare}  VQE results for a particle moving in a non-Abelian gauge field  and using the oscillator basis. All of the Hamiltonian were mapped to 9-qubit operators. The quantum circuit for each simulation utilized an \(R_y\) variational form, with a fully entangled circuit of depth 3. The backend used was a statevector simulator. The Sequential Least SQuares Programming (SLSQP) optimizer was used, with a maximum of 600 iterations.}
\end{table}

\section{Vertex operators and quantum simulation associated with non-linear field equations}

One difficulty with quantum simulation is the the quantum circuits represent linear operators as a series of Unitary gates. This works well in simulating linear equations like the Schrodinger equation. However nonlinear equations are also important in physics such as the Yang-Mills equation. To represent the nonlinear aspects of of these equations on a quantum computer one can use Vertex operators which are unitary operators and can be realized in terms of gates. This modifies the the Evolution of Hamiltonian algorithm by considering worldine actions on a graph which splits at a vertex at which point the vertex operator is inserted. This method allows the calculation of scattering amplitudes using the worldline formalism and requires far fewer qubits that a full quantum field theory simulation on a quantum computer.

For example for the $SU(2)$ Yang-Mills one needs to add an additional $x$ field in the world line action corresponding to the rank of the group and two vertex operators $V_1$ and $V_2$ associated with the roots of the Lie algebra of which there are two. For $SU(3)$ the rank is two so we add two additional $x$ fields and six vertex operators $V_i$ for $i=1,2,\dots, 6$.

\subsection{Nonlinear Klein-Gordon equation}
As a simple example consider the case of the non-linear Klein-Gordon equation \cite{Reuter:1996zm}.
The nonlinear Klein-Gordon equation with interacting potential $V(\phi)$ is:
\begin{equation}\partial _t^2\phi  - {\nabla ^2}\phi  + m^2 \phi + V'(\phi ) = 0\end{equation}
For a quartic interaction potential this equation has a solution in terms of the Jacobi elliptic functions.

The Effective action for the nonlinear Klein-Gordon field has a representation in term of a worldline path integral as:
\begin{equation}\Gamma [\phi ] = \frac{1}{2}\int_0^\infty  {\frac{{dT}}{T}} {e^{ - {m^2}T}}\int {Dx(\tau ){e^{ - \int_0^T  {d\tau \left( {\frac{1}{4}{{\dot x}^2} - V''(\phi (x(\tau )))} \right)} }}} \end{equation}
where $m$ is the mass of the Klein-Gordon field. The Hamiltonian associated with this world line action is:
\begin{equation}H = {p^\mu }{p_\mu } + {m^2} + V''(\phi )\end{equation}
For $\lambda \phi^3$ potential scattering amplitudes can be determined by inserting factors
\begin{equation}V({p_k}) = \lambda {e^{i{p_k}x({\tau _k})}}\end{equation}
into the world line path integral where $p_k$ denotes the energy-momentum of $k$th particle.
\begin{equation}{V_{scalar}}(p) = \int_0^T {d\tau {e^{ipx(\tau )}}} \end{equation}

\subsection{Nonlinear Euler-Heisenberg equation}

Another system of nonlinear equations that can be realized on a quantum computer using this method is the Non-linear Euler-Heisenberg equation \cite{Reuter:1996zm}
\cite{Bastianelli:2008vh}
\cite{Schubert:2001he}
\cite{Edwards:2019eby}
\cite{Karbstein:2019wmj}
\cite{Dunne:2004nc}
\cite{Sato:2000cr}
\cite{Huet}
\cite{Bastianelli:2008cu}
\cite{Bastianelli:2018twb}.
In the weak field limit the Euler-Heisenberg Lagrangian density is given by:
\begin{equation}L = \frac{1}{2}({E^2} - {B^2}) + \frac{{2{\alpha ^2}}}{{45{m^4}}}\left[ {{{({E^2} - {B^2})}^2} + 7{{(E \cdot B)}^2}} \right]\end{equation}
The origin of the quartic terms are quantum corrections from fermion loop diagrams.

For scalar electrodynamics the worldline representation of the Euler-Heisenberg effective action is:
\begin{equation}\Gamma [A] = \int_0^\infty  {\frac{{dT}}{T}} {e^{ - {m^2}T}}\int {Dx(\tau ){e^{ - \int_0^T  {d\tau \left( {\frac{1}{4}{{\dot x}^2} + i e{{\dot x}^\mu }{A_\mu }(x(\tau ))} \right)} }}} \end{equation}
The Hamiltonian associated with this world line action is:
\begin{equation}H = \left( {{p^\mu } + e{A^\mu }(x)} \right)\left( {{p_\mu } + e{A_\mu }} \right) + {m^2} + V''(\phi )\end{equation}
The vertex operator is:
\begin{equation}V_{scalar}^A(\varepsilon ,p) = \int_0^T {d\tau {\varepsilon _\mu }{{\dot x}^\mu }{e^{ipx(\tau )}}} \end{equation}
where $\epsilon$ is the polarization of the photon.
For spinor electrodynamics the worldline representation of the effective action is more complicated and one needs to introduce worldline Grassmann variables. It is given by:
\begin{equation}\Gamma [A] = 
- \frac{1}{2}\int_0^\infty  {\frac{{dT}}{T}} {e^{ - {m^2}T}}\int {Dx(\tau )D\psi (\tau ){e^{ - \int_0^T {d\tau \left( {\frac{1}{4}{{\dot x}^2} + \frac{1}{2}{\psi _\mu }{{\dot \psi }^\mu } + ie{{\dot x}^\mu }{A_\mu }(x(\tau )) - ie{\psi ^\mu }{F_{\mu \nu }}(x(\tau )){\psi ^\nu }} \right)} }}} \end{equation}
The vertex operator is:
\begin{equation}
    V_{spinor}^A(\varepsilon ,p) = \int_0^T {d\tau \left( {{\varepsilon _\mu }{{\dot x}^\mu } + 2i{\varepsilon _\mu }{\psi ^\mu }{k_\nu }{\psi ^\nu }} \right){e^{ip x(\tau )}}} \end{equation}

\subsection{Quantum computation with Vertex operators}

One can use quantum computation with vertex operators.  First construct a circuit from a EOH for a free particle for time $\tau_1 - \epsilon$ followed by a Unitary insertion of a Vertex operator for a time $\tau - \epsilon$ to $\tau + \epsilon$ and then finally we have a EOH from $\tau + \epsilon$ tp $T$. Schematically we have $K(0,\tau - \epsilon)V(\tau- \epsilon, \tau + \epsilon)K(\tau + \epsilon, T)$ as described in figure 5. One can also add more insertions of N vertex operators to model scattering of two scalar particles  with a nonlinear $\phi^3$ interaction \cite{smatrix}
\cite{Dai:2009vqb}
\cite{Avramis:2002xf}
\cite{Casalbuoni:1974pj}.
\begin{figure}
\centering
\minipage{0.5\textwidth}
  \includegraphics[width=\linewidth]{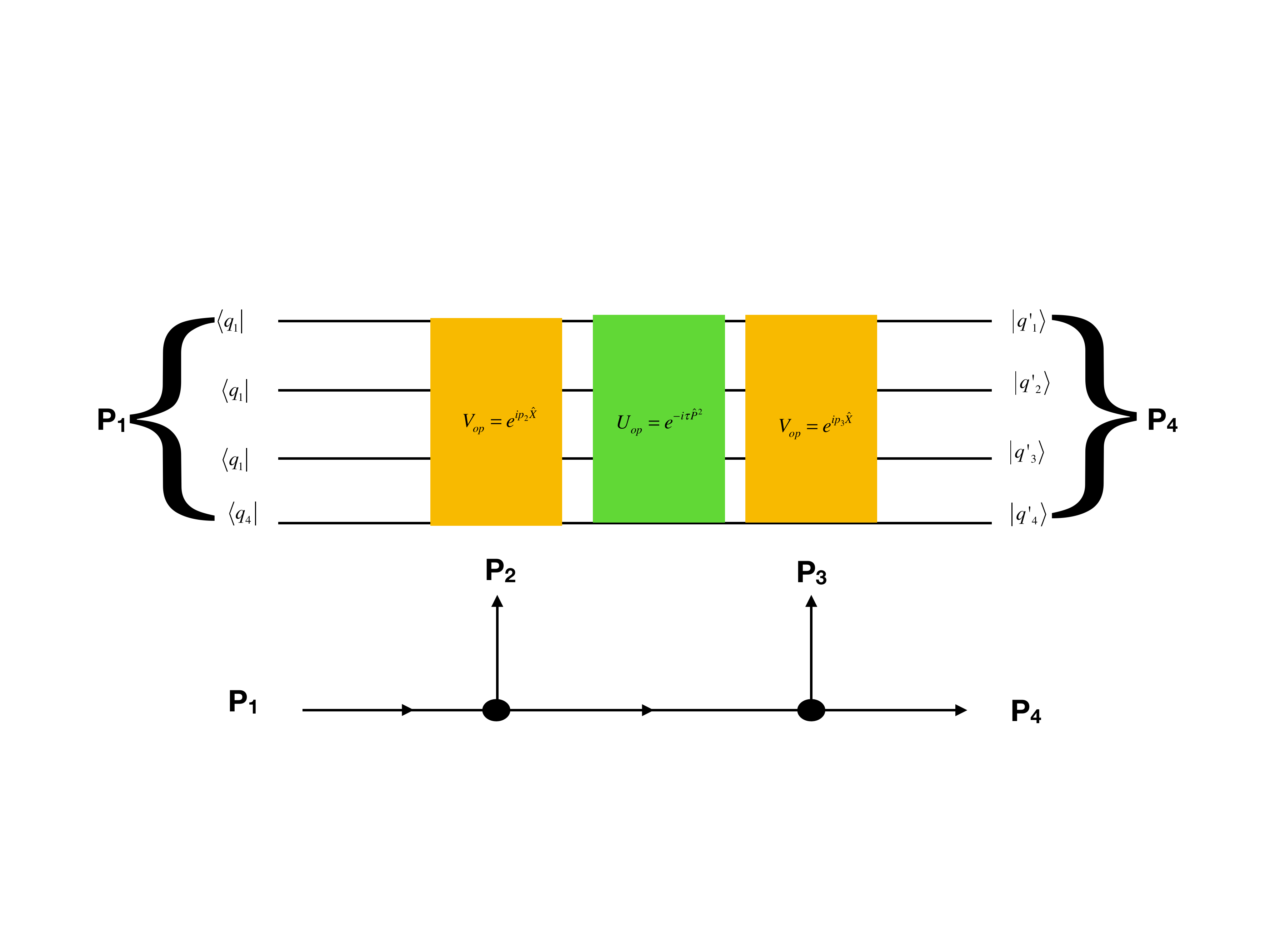}
%\label{fig:awesome_image1}
\endminipage\hfill
\minipage{0.5\textwidth}
  \includegraphics[width=\linewidth]{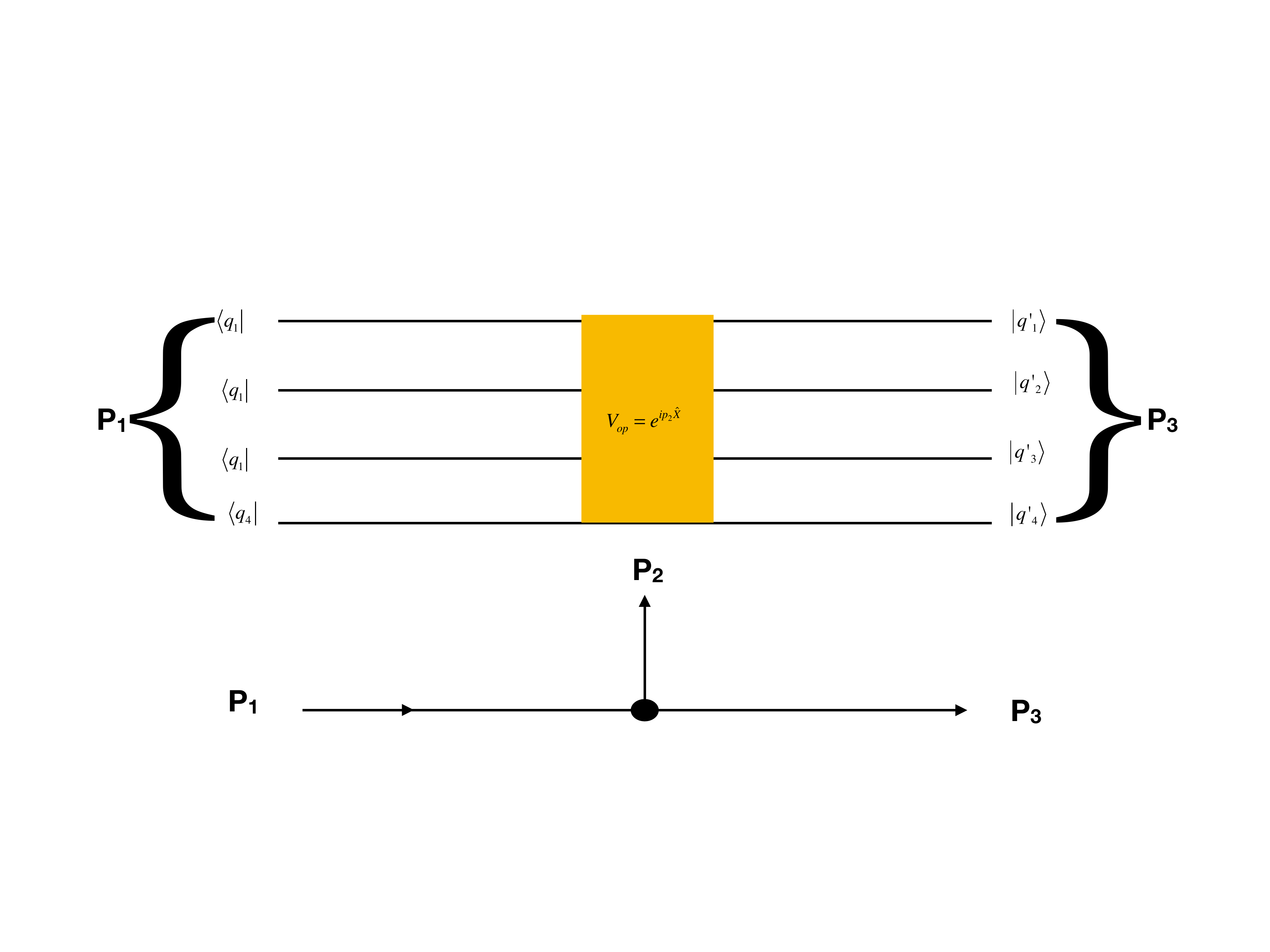}
\endminipage\hfill
\caption{(top) Scattering amplitude for scalar particle scattering in a $\phi^3$  in terms of a quantum circuit where the green rectangle indicated the unitary operator for time evolution and the orange rectangle indicates the insertion of a unitary vertex operator(bottom) Scattering amplitude in terms of a Feynman graph corresponding to the top circuit. }
\end{figure}
\begin{figure}
\centering
\minipage{0.5\textwidth}
  \includegraphics[width=\linewidth]{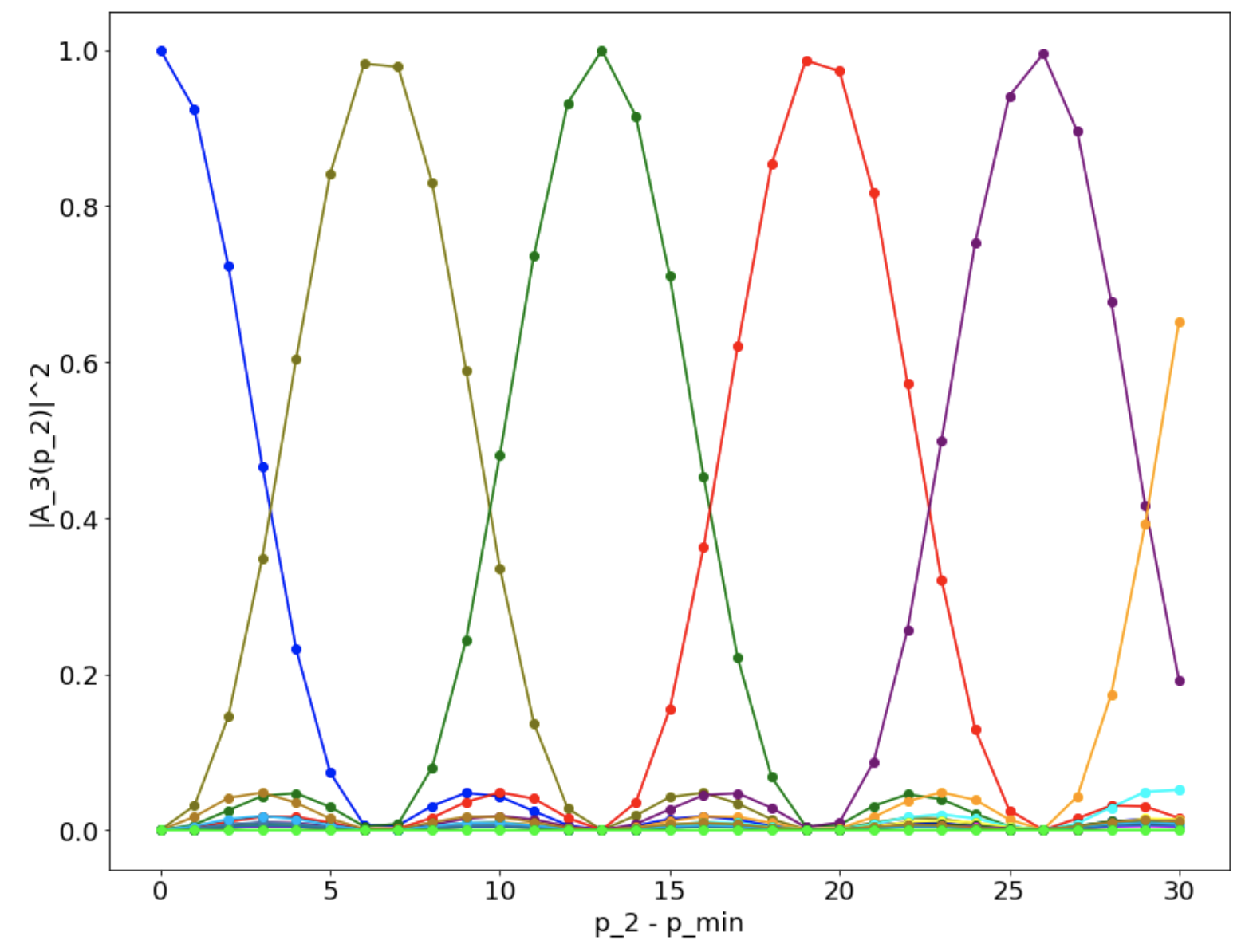}
%\label{fig:awesome_image1}
\endminipage\hfill
\minipage{0.5\textwidth}
  \includegraphics[width=\linewidth]{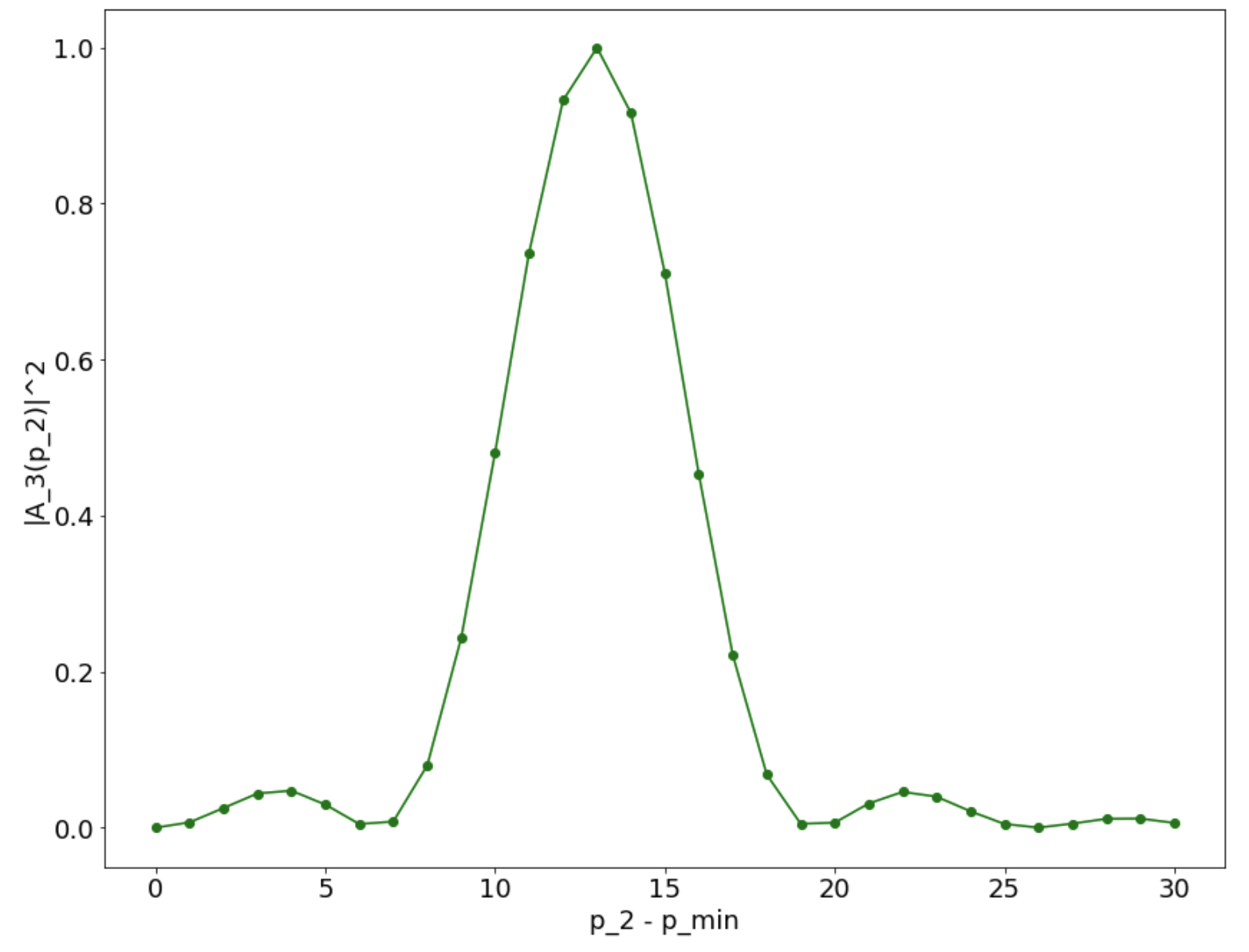}
\endminipage\hfill
\caption{(left) Result of quantum computation of a scattering amplitude for scalar particle scattering in a $\phi^3$  in terms of a quantum circuit with the insertion of a unitary vertex operator. The different graphs correspond to different momentum differences between the initial and final states after emission or absorption  of a particle. (right) Quantum computation of a scattering amplitude for scalar particle with one specific momentum difference between the initial and final states after emission or absorption  of a particle.  }
\end{figure}
For a three scalar particle amplitude given by:
\begin{equation}A({p_1},{p_2},{p_3}) = \left\langle {{p_1}} \right|{\hat{V}_{op}}({p_2})\left| {{p_3}} \right\rangle  = g{\delta ^n}\left( {{p_2} - \left( {{p_3} - {p_1}} \right)} \right)\end{equation}
one can insert the unitary vertex operator 
\begin{equation}{{\hat V}_{op}}({p_2}) = {e^{i{p_2}\hat X}}\end{equation}
into a quantum circuit and representing in the initial momentum states in terms of qubits we find the results in figure 6. Note the peak in $p_2$ as a function of the initial and final momentum difference which represents a discrete approximation to the Dirac delta function. Four qubits were used in this computation so we have sixteen possible initial momentum states in the discretization. The results of the quantum computation using the Trotter-Suzuki approximation for the insertion of the unitary vertex operator, where the evolution parameter was taken to be $p_2$. The insertion of the vertex operator leads to a $g \phi^3$ interaction potential in the effective action for $\phi$ and a quadratic term in its non-linear Klein-Gordon equation of motion.

\section{Conclusions}

In this paper we applied the VQE and EOH quantum algorithms on an IBM quantum simulator to the problem of simulating particle propagation in an magnetic field. We compared the simulation in both Cartesian and polar coordinates. We also discussed the quantum simulation of a particle in a chromo-magnetic field governed by the the non-linear Yang-Mills equation. Finally we gave example of the use of vertex operators on quantum computers to simulate the scattering in various non-linear theories and gave examples in the case of the nonlinear Klein-Gordon equation. The world line approach has several advantages over for quantum field simulation including the use of fewer qubits and being able to do simulations in 3+1 dimensions. It will interesting to explore worldline path integrals on near term quantum computers. The quantum simulation of particles in background gauge fields also has applications to Kaluza-Klein theory and quantum gravity \cite{Grosche:1988um}
\cite{Witten:1981nf}
\cite{Hosotani:1983xw}
\cite{RandjbarDaemi:1982hi}
\cite{RandjbarDaemi:1983ij}
\cite{Grinstein:1989pc} which will also be interesting to examine through quantum computing. The literature on the worldine formalism is extensive 
\cite{Langfeld:2007wh}
\cite{Gies:2001zp}
\cite{Schafer:2015wta}
\cite{Franchino-Vinas:2019udt}
\cite{Edwards:2019fjh}
\cite{Edwards:2017won}
\cite{Corradini:2020tgk}
\cite{Corradini:2016czo}
\cite{Edwards:2016acz}
 and it will be of interest to adapt these sophisticated techniques to quantum computing and explore possible advantages of quantum computing for chiral fermions, background gravitational fields and non-Abelian gauge fields. 

\section*{Acknowledgements}
This material is based upon work supported in part by the U.S. Department of Energy, Office of Science, National Quantum Information Science Research Centers, Co-design Center for Quantum Advantage (C2QA) under contract number DE-SC0012704. This project was also supported in part by the U.S. Department of Energy, Office of Science, Office of Workforce Development for Teachers and Scientists (WDTS). We thank Rob Pisarski, Raju Venugopalan and Niklas Mueller for useful discussions about worldline path integrals. We thank James P. Edwards for useful suggestions and for sending us relevant references on the worldline formalism.


\begin{thebibliography}{0}

%\cite{Grosche:1988um}

%\cite{Hosotani:1983xw}

\bibitem{Seri} M. Seri, "A walk through quantum charged particle in a (constant) magnetic field", (2007).

\bibitem{Cheng} B. Cheng, "Path-integral evaluation of the propagator for a charged particle in a constant magnetic field with the vector potential of a solenoid", Phys.Rev.A 36, 2964 (1987).

\bibitem{Feynman} R. Feynman, A. Hibbs, "Quantum mechanics and path integrals", MacGraw-Hill (1965).

\bibitem{Kandala}A. Kandala, A.Mezzacapo,  et al."Hardware-efficient variational quantum eigensolver for small molecules and quantum magnets", 2017, Nature, 549, 242. doi:10.1038/nature23879

%\cite{Pisarski:2006yk}
\bibitem{Pisarski:2006yk}
R.~D.~Pisarski,
``Fuzzy Bags and Wilson Lines,''
Prog. Theor. Phys. Suppl. \textbf{168}, 276-284 (2007)
doi:10.1143/PTPS.168.276
[arXiv:hep-ph/0612191 [hep-ph]].

%\cite{KorthalsAltes:1994xx}
\bibitem{KorthalsAltes:1994xx}
C.~Korthals Altes, K.~Y.~Lee and R.~D.~Pisarski,
``Effective potential for the Wilson line in the standard model,''


%\cite{Mueller:2017wom}
\bibitem{Mueller:2017wom}
N.~Mueller and R.~Venugopalan,
``World-line approach to chiral kinetic theory in topological background gauge fields,''
PoS \textbf{CPOD2017}, 047 (2018)
doi:10.22323/1.311.0047
[arXiv:1712.04057 [hep-ph]].

%\cite{Mueller:2020vha}
%\cite{Callan:1989nz}
%\cite{Fan:2017uqy}
%\cite{Sen:1985eb}
%\cite{deAlwis:1986rw}
%\cite{Ellwanger:1988cc}
\bibitem{Mueller:2020vha}
N.~Mueller, A.~Tarasov and R.~Venugopalan,
``Computing real time correlation functions on a hybrid classical/quantum computer,''
Nucl. Phys. A \textbf{1005}, 121889 (2021)
doi:10.1016/j.nuclphysa.2020.121889
[arXiv:2001.11145 [hep-th]].

%\cite{Callan:1989nz}
\bibitem{Callan:1989nz}
C.~G.~Callan, Jr. and L.~Thorlacius,
``SIGMA MODELS AND STRING THEORY,''
Print-89-0232 (PRINCETON).
%3 citations counted in INSPIRE as of 04 Apr 2021

%\cite{Fan:2017uqy}
\bibitem{Fan:2017uqy}
W.~Fan, A.~Fotopoulos, S.~Stieberger and T.~R.~Taylor,
``SV-map between Type I and Heterotic Sigma Models,''
Nucl. Phys. B \textbf{930}, 195-218 (2018)
doi:10.1016/j.nuclphysb.2018.02.024
[arXiv:1711.05821 [hep-th]].

%\cite{Sen:1985eb}
\bibitem{Sen:1985eb}
A.~Sen,
``The Heterotic String in Arbitrary Background Field,''
Phys. Rev. D \textbf{32}, 2102 (1985)
doi:10.1103/PhysRevD.32.2102

%\cite{deAlwis:1986rw}
\bibitem{deAlwis:1986rw}
S.~P.~de Alwis,
``Strings in Background Fields, Beta Functions and Vertex Operators,''
Phys. Rev. D \textbf{34}, 3760 (1986)
doi:10.1103/PhysRevD.34.3760

%\cite{Ellwanger:1988cc}
\bibitem{Ellwanger:1988cc}
U.~Ellwanger, J.~Fuchs and M.~G.~Schmidt,
``The Heterotic $\sigma$ Model With Background Gauge Fields,''
Nucl. Phys. B \textbf{314}, 175 (1989)
doi:10.1016/0550-3213(89)90117-X

%\cite{Marinho:2009tm}
%\cite{Norfjand:2019gof}
%\cite{Marciano:1977sm}
%\cite{Actor:1979in}
%\cite{Moody:1985ty}
\bibitem{Marinho:2009tm}
J.~A.~O.~Marinho, O.~Oliveira, B.~V.~Carlson and T.~Frederico,
``Revisiting the Wu-Yang Monopole: Classical solutions and conformal invariance,''
[arXiv:0912.3274 [hep-th]].

%\cite{Norfjand:2019gof}
\bibitem{Norfjand:2019gof}
F.~N\o{}rfjand and N.~T.~Zinner,
``Non-existence theorems and solutions of the Wu-Yang monopole equation,''
[arXiv:1911.08140 [hep-th]].

%\cite{Marciano:1977sm}
\bibitem{Marciano:1977sm}
W.~J.~Marciano,
``Magnetic Monopoles and Nonabelian Gauge Theories,''
Int. J. Theor. Phys. \textbf{17}, 275 (1978)
doi:10.1007/BF00672873

%\cite{Actor:1979in}
\bibitem{Actor:1979in}
A.~Actor,
``Classical Solutions of SU(2) Yang-Mills Theories,''
Rev. Mod. Phys. \textbf{51}, 461 (1979)
doi:10.1103/RevModPhys.51.461

%\cite{Moody:1985ty}
\bibitem{Moody:1985ty}
J.~Moody, A.~D.~Shapere and F.~Wilczek,
``Simple Realizations of Magnetic Monopole Gauge Fields: Diatoms and Spin Precession,''
Phys. Rev. Lett. \textbf{56}, 893 (1986)
doi:10.1103/PhysRevLett.56.893

%\cite{Reuter:1996zm}
%\cite{Bastianelli:2008vh}
%\cite{Schubert:2001he}
%\cite{Edwards:2019eby}
%\cite{Karbstein:2019wmj}
%\cite{Dunne:2004nc}
%\cite{Sato:2000cr}
%\cite{Huet}
%\cite{Bastianelli:2008cu}
%\cite{Bastianelli:2018twb}

\bibitem{Reuter:1996zm}
M.~Reuter, M.~G.~Schmidt and C.~Schubert,
``Constant external fields in gauge theory and the spin 0, 1/2, 1 path integrals,''
Annals Phys. \textbf{259}, 313-365 (1997)
doi:10.1006/aphy.1997.5716
[arXiv:hep-th/9610191 [hep-th]].

\bibitem{Bastianelli:2008vh}
F.~Bastianelli, O.~Corradini, P.~A.~Pisani and C.~Schubert,
``Scalar heat kernel with boundary in the worldline formalism,''
JHEP \textbf{10}, 095 (2008)
[arXiv:0809.0652 [hep-th]].

%\cite{Schubert:2001he}
\bibitem{Schubert:2001he}
C.~Schubert,
``Perturbative quantum field theory in the string inspired formalism,''
Phys. Rept. \textbf{355}, 73-234 (2001)
[arXiv:hep-th/0101036 [hep-th]].

\cite{Edwards:2019eby}
\bibitem{Edwards:2019eby}
J.~P.~Edwards and C.~Schubert,
``Quantum mechanical path integrals in the first quantised approach to quantum field theory,''
[arXiv:1912.10004 [hep-th]].

%\cite{Karbstein:2019wmj}
\bibitem{Karbstein:2019wmj}
F.~Karbstein,``All-Loop Result for the Strong Magnetic Field Limit of the Heisenberg-Euler Effective Lagrangian,''
Phys. Rev. Lett. \textbf{122}, no.21, 211602 (2019)
[arXiv:1903.06998 [hep-th]].

%\cite{Dunne:2004nc}
\bibitem{Dunne:2004nc}
G.~V.~Dunne,
``Heisenberg-Euler effective Lagrangians: Basics and extensions,''
[arXiv:hep-th/0406216 [hep-th]].

%\cite{Sato:2000cr}
\bibitem{Sato:2000cr}
H.~T.~Sato, M.~G.~Schmidt and C.~Zahlten,
``Two loop Yang-Mills theory in the worldline formalism and an Euler-Heisenberg type action,''
Nucl. Phys. B \textbf{579}, 492-524 (2000)
[arXiv:hep-th/0003070 [hep-th]].

\bibitem{Huet} Huet, I., de Traubenberg, M.R.  Schubert, C. Three-loop Euler-Heisenberg Lagrangian in 1+1 QED. Part I. Single fermion-loop part. J. High Energ. Phys. 2019, 167 (2019).

%\cite{Bastianelli:2008cu}
\bibitem{Bastianelli:2008cu}
F.~Bastianelli, J.~M.~Davila and C.~Schubert,
``Gravitational corrections to the Euler-Heisenberg Lagrangian,''
JHEP \textbf{03}, 086 (2009)
[arXiv:0812.4849 [hep-th]].

%\cite{Bastianelli:2018twb}
\bibitem{Bastianelli:2018twb}
F.~Bastianelli, O.~Corradini and L.~Iacconi,
``Simplified path integral for supersymmetric quantum mechanics and type-A trace anomalies,''
JHEP \textbf{05}, 010 (2018)
[arXiv:1802.05989 [hep-th]].









%\cite{smatrix}
%\cite{Dai:2009vqb}
%\cite{Avramis:2002xf}
%\cite{Casalbuoni:1974pj}
\bibitem{smatrix} Tevzadze, Revaz and Gia Giorgadze. “Quantum computation with scattering matrices.” Journal of Mathematical Sciences 153 (2006): 197-209.

%\cite{Dai:2009vqb}
\bibitem{Dai:2009vqb}
P.~Dai,
``Worldgraph approach to amplitudes,''

%\cite{Avramis:2002xf}
\bibitem{Avramis:2002xf}
S.~D.~Avramis, A.~I.~Karanikas and C.~N.~Ktorides,
``Perturbative computation of the gluonic effective action via Polyakov's worldline path integral,''
Phys. Rev. D \textbf{66}, 045017 (2002)
doi:10.1103/PhysRevD.66.045017
[arXiv:hep-th/0205272 [hep-th]].

%\cite{Casalbuoni:1974pj}
\bibitem{Casalbuoni:1974pj}
R.~Casalbuoni, J.~Gomis and G.~Longhi,
``The Relativistic Point Revisited in the Light of the String Model,''
Nuovo Cim. A \textbf{24}, 249 (1974)
doi:10.1007/BF02821992

%\cite{Grosche:1988um}
%\cite{Witten:1981nf}
%\cite{Hosotani:1983xw}
%\cite{RandjbarDaemi:1982hi}
%\cite{RandjbarDaemi:1983ij}
%\cite{Grinstein:1989pc}
\bibitem{Grosche:1988um}
C.~Grosche,
``The Path Integral on the Poincare Upper Half Plane With a Magnetic Field and for the Morse Potential,''
Annals Phys. \textbf{187}, 110 (1988)
doi:10.1016/0003-4916(88)90283-7

%\cite{Witten:1981nf}
\bibitem{Witten:1981nf}
E.~Witten,
``Dynamical Breaking of Supersymmetry,''
Nucl. Phys. B \textbf{188}, 513 (1981)
doi:10.1016/0550-3213(81)90006-7

\bibitem{Hosotani:1983xw}
Y.~Hosotani,
``Dynamical Mass Generation by Compact Extra Dimensions,''
Phys. Lett. B \textbf{126}, 309-313 (1983)
doi:10.1016/0370-2693(83)90170-3


%\cite{RandjbarDaemi:1982hi}
\bibitem{RandjbarDaemi:1982hi}
S.~Randjbar-Daemi, A.~Salam and J.~A.~Strathdee,
``Spontaneous Compactification in Six-Dimensional Einstein-Maxwell Theory,''
Nucl. Phys. B \textbf{214}, 491-512 (1983)
doi:10.1016/0550-3213(83)90247-X

%\cite{RandjbarDaemi:1983ij}
\bibitem{RandjbarDaemi:1983ij}
S.~Randjbar-Daemi, A.~Salam and J.~A.~Strathdee,
``Stability of Instanton Induced Compactification in Eight-dimensions,''
Nucl. Phys. B \textbf{242}, 447-472 (1984)
doi:10.1016/0550-3213(84)90404-8


%\cite{Grinstein:1989pc}
\bibitem{Grinstein:1989pc}
B.~Grinstein and J.~Maharana,
``Vertex Operators for Axionic Wormholes,''
Nucl. Phys. B \textbf{333}, 160-172 (1990)
doi:10.1016/0550-3213(90)90226-4

%\cite{Gies:2001zp}
%\cite{Schafer:2015wta}
%\cite{Franchino-Vinas:2019udt}
%\cite{Edwards:2019fjh}
%\cite{Edwards:2017won}
%\cite{Corradini:2020tgk}
%\cite{Corradini:2016czo}
%\cite{Langfeld:2007wh}
\bibitem{Langfeld:2007wh}
K.~Langfeld, G.~Dunne, H.~Gies and K.~Klingmuller,
%``Worldline Approach to Chiral Fermions,''
PoS \textbf{LATTICE2007}, 202 (2007)
doi:10.22323/1.042.0202
[arXiv:0709.4595 [hep-lat]].
\bibitem{Gies:2001zp}

H.~Gies and K.~Langfeld,
``Quantum diffusion of magnetic fields in a numerical worldline approach,''
Nucl. Phys. B \textbf{613}, 353-365 (2001)
doi:10.1016/S0550-3213(01)00377-7
[arXiv:hep-ph/0102185 [hep-ph]].

%\cite{Schafer:2015wta}
\bibitem{Schafer:2015wta}
M.~Schafer, I.~Huet and H.~Gies,
``Worldline Numerics for Energy-Momentum Tensors in Casimir Geometries,''
J. Phys. A \textbf{49}, no.13, 135402 (2016)
doi:10.1088/1751-8113/49/13/135402
[arXiv:1509.03509 [hep-th]].

%\cite{Franchino-Vinas:2019udt}
\bibitem{Franchino-Vinas:2019udt}
S.~Franchino-Vi\~nas and H.~Gies,
``Propagator from Nonperturbative Worldline Dynamics,''
Phys. Rev. D \textbf{100}, no.10, 105020 (2019)
doi:10.1103/PhysRevD.100.105020
[arXiv:1908.04532 [hep-th]].

%\cite{Edwards:2019fjh}
\bibitem{Edwards:2019fjh}
J.~P.~Edwards, U.~Gerber, C.~Schubert, M.~A.~Trejo, T.~Tsiftsi and A.~Weber,
``Applications of the worldline Monte Carlo formalism in quantum mechanics,''
Annals Phys. \textbf{411}, 167966 (2019)
doi:10.1016/j.aop.2019.167966
[arXiv:1903.00536 [quant-ph]].

%\cite{Edwards:2017won}
\bibitem{Edwards:2017won}
J.~P.~Edwards, U.~Gerber, C.~Schubert, M.~A.~Trejo and A.~Weber,
``Integral transforms of the quantum mechanical path integral: hit function and path averaged potential,''
Phys. Rev. E \textbf{97}, no.4, 042114 (2018)
doi:10.1103/PhysRevE.97.042114
[arXiv:1709.04984 [quant-ph]].

%\cite{Corradini:2020tgk}
\bibitem{Corradini:2020tgk}
O.~Corradini and M.~Muratori,
``A Monte Carlo Approach to the Worldline Formalism in Curved Space,''
JHEP \textbf{11}, 169 (2020)
doi:10.1007/JHEP11(2020)169
[arXiv:2006.02911 [hep-th]].

%\cite{Corradini:2016czo}
\bibitem{Corradini:2016czo}
O.~Corradini and J.~P.~Edwards,
``Mixed symmetry tensors in the worldline formalism,''
JHEP \textbf{05}, 056 (2016)
doi:10.1007/JHEP05(2016)056
[arXiv:1603.07929 [hep-th]].

%\cite{Edwards:2016acz}
\bibitem{Edwards:2016acz}
J.~P.~Edwards and O.~Corradini,
``Mixed symmetry Wilson-loop interactions in the worldline formalism,''
JHEP \textbf{09}, 081 (2016)
doi:10.1007/JHEP09(2016)081
[arXiv:1607.04230 [hep-th]].

\end{thebibliography}
\end{document}